\def\wide#1#2{
\end{multicols}
\vspace{-0.1in} \widetext \noindent \if#1t \else
\raisebox{9pt}[0in][0.0in]
{$\rule{3.4in}{0.4pt}\rule{0.4pt}{6pt}$\hspace{3.6in}} \fi
\vspace{-0.15in} #2 \vspace{-0.2in} \if#1b \else
\raisebox{-9pt}[0in][0.0in]
{\hspace{3.55in}$\rule{0.4pt}{6pt}\rule[6pt]{3.5in}{0.4pt}$} \fi
\begin{multicols}{2}
\vspace{-0.05in} \noindent
}
\documentstyle[aps,prb,multicol,epsf]{revtex}
\begin{document}

\author{A.\ E.\ Koshelev and V.\ M.\ Vinokur}
\address{Materials Science Division, Argonne National Laboratory,
Argonne, Illinois 60439}
\title{Suppression of surface barrier in superconductors by columnar
defects}
\date{\today}
\maketitle
\begin{abstract}
We investigate the influence of columnar defects in layered
superconductors on the thermally activated penetration of pancake
vortices through the surface barrier.  Columnar defects, located
near the surface, facilitate penetration of vortices through the
surface barrier, by creating ``weak spots'', through which
pancakes can penetrate into the superconductor.  Penetration of a
pancake mediated by an isolated column, located near the surface,
is a two-stage process involving hopping from the surface to the
column and the detachment from the column into the bulk; each
stage is controlled by its own activation barrier.  The resulting
effective energy is equal to the maximum of those two barriers.
For a given external field there exists an optimum location of the
column for which the barriers for the both processes are equal and
the reduction of the effective penetration barrier is maximal. At
high fields the effective penetration field is approximately two
times smaller than in unirradiated samples.  We also estimate the
suppression of the effective penetration field by column clusters.
This mechanism provides further reduction of the penetration field
at low temperatures.
\end{abstract}
\begin{multicols}{2}
\section{Introduction}

The properties of the Abrikosov vortex state of type II
superconductors with artificially manufactured columnar defects
attract a great deal of current attention.  Motivated originally
by a technological quest for enhanced vortex pinning,
superconductors with columnar defects revealed a vast diversity
of remarkable phenomena. The possibility to introduce controlled
disorder and to tune parameters (such as vortex density and
interactions between vortices) has made them one of the favorite
experimental systems for studies of the statistical mechanics and
dynamics of a glassy state (see recent review article
\onlinecite{BeekPRB00} and references therein).

On the other hand, the magnetic response of high-temperature
superconductors is known to be controlled to a large extent by the
creep of vortices over the Bean-Livingston surface barrier,
\cite{Kopylov90,Konczykowski91,Mints93,Zeldov94,Burlachkov94} an
important manifestation of which is the exponential temperature
dependence of the effective penetration field.  It was recently
observed that columnar defects can strongly facilitate the creep
of pancake vortices over the surface barrier and reduce the
penetration field.\cite{Gregory01} Usually pinning by disorder and
surface barrier are considered to be competing effects that
alternatively control the magnetic response in the vortex state.
In this paper we consider an interplay between surface and bulk
pinning and develop a theory for the disorder-assisted surface
creep
in highly anisotropic superconductors focusing on the case of
randomly distributed columnar defects.  This effect is somewhat
analogous to tunnelling of quantum particle in the presence of
sub-barrier disorder.\cite{LifshitzJETP79}

\begin{figure}
\epsfxsize=2.3in \epsffile{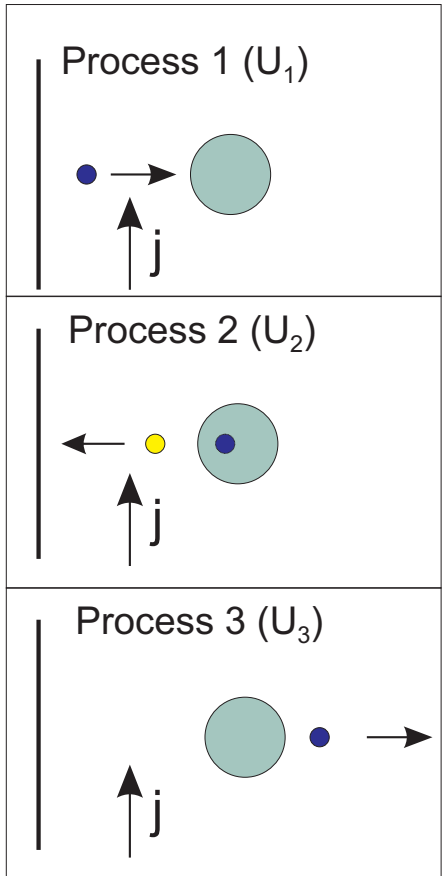} \caption{Mechanisms of
penetration of pancake vortex from the surface into the bulk of a
superconductor:(a)Nucleation of the vortex at the column via the
motion of vortex from the surface (b)Nucleation of vortex at the
column via motion of an antivortex from the column to the surface
(c) motion of the vortex from the column into the bulk}
\label{Fig-Processes}
\end{figure}

It is clear from a general consideration that surface
imperfections create weak spots, facilitating penetration of
vortices through the Bean-Livingston barrier. However, a
quantitative theory for imperfection of arbitrary kind is not
available. Special kinds of surface irregularities have been
considered in Refs.\
\onlinecite{BuzdPhysC98,MelnPhysC01,VodolasovPRB00}. We address
well defined surface disorder created by controlled irradiation
with heavy ions. In this case vortices enter the sample near the
weak spots where columnar defects are located close enough to the
surface to suppress the surface barrier.  Vortex penetration
consists of two steps (see Fig.\ \ref{Fig-Processes}): (i) the
capturing of the vortex onto a near-surface column and (ii) the
detachment of the trapped vortex into the bulk .  The resulting
effective barrier for pancake penetration via an isolated column
is the maximal value of the two barriers corresponding to the
above processes. The capturing process has, in its turn, a
two-channel character and may occur either via the direct motion
of a pancake to a column or via the nucleation of an {\it
antivortex} at the column and its subsequent advance towards the
surface (see Fig.\ \ref{Fig-Processes}).  The capturing process is
controlled by the channel with the {\it smallest} barrier.  If the
mediating column is located far from the surface, the penetration
process is controlled by the transfer of a vortex from the surface
to this column.  When the column is sufficiently close to the
surface, the controlling barrier corresponds to detaching a vortex
from the column to the bulk. For every external field there exists
an optimal location of the column for which the reduction of the
barrier is maximal, and penetration of pancakes into the sample
occurs mainly via such optimally placed columns.

One can expect that the surface barrier may be suppressed even more
efficiently in very rare places where several columns happen to be near
the surface.  The net contribution to the penetration rate from such
events is determined as a balance between their small probability and
the strong local suppression of the barrier.  We estimate in section
\ref{Sec-Coll} the collective suppression by the column clusters.  At
low temperatures the collective suppression always becomes more
efficient than the single-column mechanism.

\section{Vortex energy near surface in presence of an isolated
columnar defect}

\begin{figure}
\epsfxsize=3.2in \epsffile{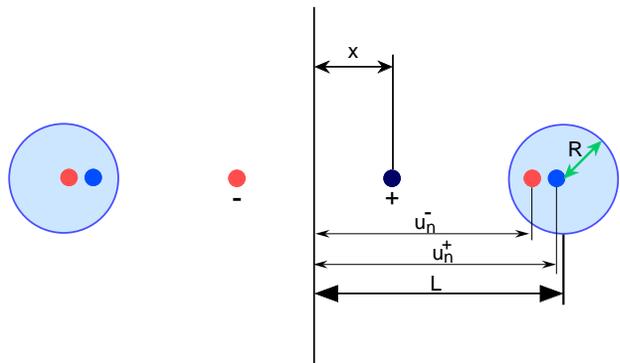} \caption{Geometry: a
pancake vortex between the surface and a columnar defect.
Interaction with the surface and the column can be described in
terms of an infinite set of positive and negative images obtained
by multiple subsequent reflections by the surface and the column
(only the first two images inside the column are shown).}
\label{Fig-Images}
\end{figure}

We consider an irradiated superconductor with insulating columnar
defects oriented along the c-axis in external magnetic field also
applied along the c-axis.  Let an isolated columnar defect have
radius $R$ and be located at a distance $L$ from the surface of
superconductor (see Fig.\ \ref{Fig-Images}).  The energy of a
pancake vortex located between the surface and column at distance
$x$ from the surface consists of two parts: the direct
interactions with the column and the surface, $U_{int}(x)$, and
the interaction with the Meissner current, $ U_{j}(x)$.  We
introduce dimensionless variables measuring length in units of
$R$, energy in units of $s\varepsilon _{0}\equiv s\Phi_0^2/(4\pi
\lambda)^2$, current in units of $ \frac{c\Phi _{0}}{(4\pi \lambda
)^{2}R}$, and magnetic field in units of $\frac{\Phi _{0}}{4\pi
\lambda R}$. Here $\lambda$ is the in-plane London penetration
depth and $s$ is the interlayer spacing. We also use the notation
$l\equiv L/R$.

\subsection{Direct interaction with surface and column}

The interaction of the vortex with either the plain surface or
with an isolated column can be calculated by virtue of the image
technique.\cite{Image-Note}  The interaction with the surface is
obtained by placing a negative vortex at the point $-x$ and the
interaction with the isolated column can be obtained by placing a
negative vortex at point $l-\frac{1}{l-x}$ inside the column and a
positive vortex at the column center.\cite{Vort-Column}  In the
case where the vortex is confined between the column cavity and
the surface, adding these images only would not solve the problem
because the currents due to the surface image do not satisfy the
boundary conditions at the surface of the columnar cavity and vice
versa. To compensate the currents of the surface image, one has to
add the its image (reflection) inside the column.  This eliminates
vortex in the column center  and adds a vortex at the point
$l-\frac{1}{l+x}$. After that we have to add the surface image of
this vortex at the point $-l+\frac{1}{l+x}$.  Continuing these
reflections, we obtain an infinite set of positive and negative
images.  We label coordinates of the positive (negative) images
inside the column after the $n$-th double reflection as
$u_{n}^{+}$ ($u_{n}^{-}$) (every such image has a corresponding
surface image of opposite sign at $-u_{n}^{+}$($-u_{n}^{-}$)). The
$ n+1$-st image is obtained by reflecting the $n$-th image with
respect to the surface, and then reflecting it again with respect
to the column. This yields the following recurrency relations
\begin{equation}
u_{n+1}^{-}=l-\frac{1}{l+u_{n}^{-}};\;u_{n+1}^{+}=l-\frac{1}{l+u_{n}^{+}},
\label{RecurRel}
\end{equation}
with
\[
u_{1}^{-}=l-\frac{1}{l-x};\;u_{1}^{+}=l-\frac{1}{l+x},
\]
As follows from (\ref{RecurRel}), both the negative and positive
images approach the same limiting position $u_{\infty
}=\sqrt{l^{2}-1}$ as $ n\rightarrow \infty $. The interaction
energy is then expressed as an infinite series in the coordinates
of all images:
\begin{eqnarray}
U_{int}(x)&=&\ln \frac{1.47R}{\xi }+\ln 2x \label{IntEner}\\
&+&\sum_{n=1}^{\infty }\ln \left|
\frac{\left( u_{n}^{-}-x\right) \left( u_{n}^{+}+x\right) }{\left(
u_{n}^{-}+x\right) \left( u_{n}^{+}-x\right) }\right| \nonumber .
\end{eqnarray}
Note that this expression is valid for a vortex located at either
side of the column, i.e., for both $0<x<l-1$ and $x>l+1$. The
supercurrent distribution around the column containing the
trapped vortex coincides with the distribution corresponding to a
vortex placed at the point $\sqrt{ l^{2}-1}$ and its surface
image.  The energy of this state is given by
\[
U_{tr}(l)=\ln (l+\sqrt{l^{2}-1})
\]
The energy, corresponding to an antivortex at the point $x$, which
controls the competing process of antivortex motion  from the
column to the surface (we remind that in this process a
vortex-antivortex pair nucleates at the column, then the {\it
antivortex} leaves the column and hops to the surface, while the
{\it vortex} remains trapped in the column) is obtained
analogously and reads
\begin{equation}
U_{int}^{(av)}(x)=U_{int}(x)+2\ln
\frac{\sqrt{l^{2}-1}-x}{\sqrt{l^{2}-1}+x} +\ln
(l+\sqrt{l^{2}-1}).  \label{IntEnAV}
\end{equation}
Here the second term in the RHS describes the interaction of
antivortex with the trapped vortex while the third term gives the
self-energy of the trapped vortex.

To solve the recurrency relations (\ref{RecurRel}), we introduce a
new variable $f_{n}$ for both the positive and negative images
$u_{n}\equiv u_{n}^{\pm }$ as
\[
f_{n}=-\frac{\sqrt{l^{2}-1}+u_{n}}{\sqrt{l^{2}-1}-u_{n}}
\]
and transform Eq.\ (\ref{RecurRel}) into a simple relation
\[
f_{n+1}=a^{2}f_{n}
\]
with $a=l+\sqrt{l^{2}-1}>1$, which can be easily solved:
$f_{n}=a^{2n}f_{1}$.  This allows us to obtain closed analytical
formulas for the image coordinates:
\begin{eqnarray}
u_{n}^{+} &=&\sinh b\frac{x+\tanh (nb)\sinh b}{\tanh (nb)x+\sinh b}
\label{ImageSolution} \\
u_{n}^{-} &=&\sinh b\frac{x-\tanh (nb)\sinh b}{\tanh (nb)x-\sinh b}
\nonumber
\end{eqnarray}
with $b\equiv \ln \left( l+\sqrt{l^{2}-1}\right) $ and $l=\cosh
b$. Taken together, Eqs.(\ref{IntEner}) and (\ref{ImageSolution})
solve the problem of finding the  energy of a vortex located on
the perpendicular to the surface passing through the center of the
column. This result can also be obtained using complex plane
representation and the conformal mapping
\[
w=\ln \left [ \frac{\sqrt{l^{2}-1}+z}{\sqrt{l^{2}-1}-z}\right].
\]
It transforms the semispace $x>0$ with a circular hole at
$z=(l,0)$ into the rectangular area $\left\{0< w_{1} < b, -\pi <
w_{2}< \pi\right\}$ with the periodical boundary condition along
the $w_{2}$ direction.

\subsection{Interaction with the Meissner currents}

\begin{figure}
\epsfxsize=3.1in \epsffile{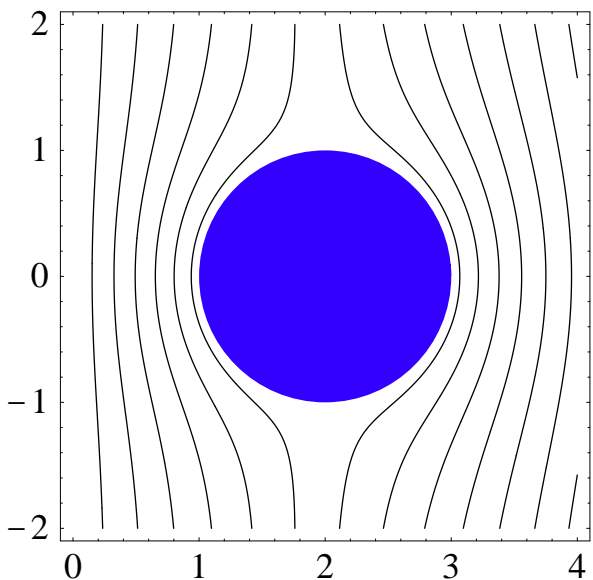} \caption{Example of the
current distribution for $l=2$. } \label{Fig-CurDist}
\end{figure}

A column (cylindrical cavity) placed near the surface disturbs
the pattern of the screening supercurrent induced by the external
magnetic field and changes accordingly the contribution to the
vortex energy arising from surface screening current.  The current
${\bf j}({\bf r})$ has to satisfy ${\rm div}{\bf j}=0$. This means
that it can be expressed in terms of a supercurrent potential,
$\phi _{j}({\bf r})$, as
\[
j_{y}=-\frac{\partial \phi _{j}}{\partial x},\;j_{x}=\frac{\partial
\phi _{j}
}{\partial y}.
\]
We consider the situation where all relevant distances are smaller
than the London penetration depth so that we can neglect screening
effects and assume ${\rm curl}{\bf j}=0$,  which implies that the
potential satisfies the Laplace equation $\Delta \phi _{j}=0$. The
problem of finding the current distribution is thus equivalent to
the problem of the polarization of a metallic cylinder near a
metallic surface by the external electric field parallel to the
surface.\cite{NoteAnalogy}  Since the normal components of ${\bf
j}$ should vanish at the surface {\it and} at the column, both the
external boundary of the superconductor and the boundary of the
column should be equipotential surfaces. Setting $\phi
_{j}(0,y)=0$, we define $\phi _{j}({\bf r})$ as the interaction
energy of a vortex at the point $ {\bf r}$ with the Meissner
current. Note that at large distances from the column, the
$x$-component of the current should vanish, thus, far from the
column ${\bf j}=(0,j)$ with $ j=cH/\left( 4\pi \lambda \right) $.
The current distribution near an isolated cylinder can be obtained
by putting vortex dipole $(j,\;0)$ at the center of the cylinder,
this dipole induces potential $\phi _{j}=-jx\left(
1-\frac{1}{(x-L)^{2}+y^{2}}\right) $.  To satisfy the boundary
condition at the surface one has to add surface image of this
dipole, i.e. to put dipole $(-j,\;0)$ at $x=-L$. However, the
currents of this surface image do not satisfy the boundary
conditions for the column and we again have to add the column
image of the surface image. Continuing this process, we again
obtain an infinite set of dipole images inside the column. Note
that this sequential reflection in the column does not preserve
the magnitude of a dipole.  The reflecting pair of opposite
vortices located near the point $(x,0)$, we derive that the
magnitude of column reflection of dipole is smaller by factor
$1/\left( l-x\right) ^{2}$ than the magnitude of the original
dipole. Denoting the coordinate of the dipole resulting from the
$n$ double reflections as $x_{n}$ and its magnitude as $jp_{n}$,
we derive the recurrency relations
\begin{equation}
p_{n+1}=\frac{1}{\left( l+x_{n}\right) ^{2}}p_{n},\;x_{n+1}=l-\frac{1}{
l+x_{n}}, \label{RecRelDip}
\end{equation}
with $x_{1}=l,\;p_{1}=1$. Analytical solutions to these equations are
given
by
\begin{eqnarray}
x_{n} &=&\sqrt{l^{2}-1}\coth nb,  \nonumber \\
p_{n} &=&\frac{l^{2}-1}{\sinh ^{2}nb}.
\end{eqnarray}
where, again, $b\equiv \ln \left( l+\sqrt{l^{2}-1}\right) $. The
potential
can be represented as an infinite series
\[
\phi _{j}({\bf r})=-jx+j\sum_{n=1}^{\infty }\left[ \frac{\left(
x-x_{n}\right) p_{n}}{\left( x-x_{n}\right) ^{2}+y^{2}}+\frac{\left(
x+x_{n}\right) p_{n}}{\left( x+x_{n}\right) ^{2}+y^{2}}\right].
\]
The interaction energy of the vortex, located at the line $y=0$, with
the Meissner current is given by
\[
\phi _{j}(x)\equiv \phi _{j}(x,0)=-jx+2xj\sum_{n=1}^{\infty
}\frac{p_{n}}{
x^{2}-x_{n}^{2}}.
\]
Fig.\ \ref{Fig-CurDist}  illustrates the current distribution for
$l=2$.  When column is located sufficiently close to the surface,
$l\rightarrow 1$, the current at the surface diverges as
$j_{y}(0,0)= \sqrt{2}j/\sqrt{l-1}$, while the current at the
opposite side of the column approaches a universal value,
$j_{y}(l+1,0)\rightarrow j\left( 1+2\sum_{n=1}^{\infty
}\frac{4n^{2}+1}{\left( 4n^{2}-1\right) ^{2}}\right)=(\pi ^2/4)j
\approx 2.47j$.

The vortex-Meissner current interaction energy becomes $\phi
_{j}(l-1)=-j \sqrt{l^{2}-1}$, when the vortex is trapped by the
column,  and the total energy of the trapped vortex reads:
\[
U_{tr}(l,j)=\ln (l+\sqrt{l^{2}-1})-j\sqrt{l^{2}-1}.
\]

\subsection{Activation barriers}

\begin{figure}
\epsfxsize=3.2in \epsffile{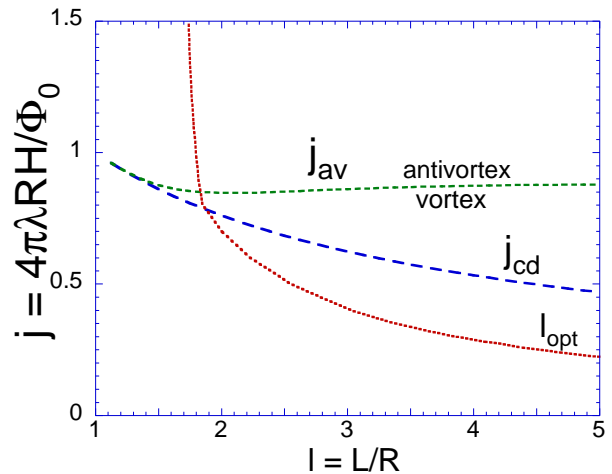} \caption{ Lines in the $j$-$l$
plane at which the penetration mechanism changes: at currents
$j>j_{cd}(l)$ it is energetically favorable to put one flux
quantum on the column, at currents $j>j_{av}(l)$ nucleation of the
flux quantum at the column occurs by motion of an antivortex from
the column to the surface, the line $l_{opt}(j)$ gives the optimum
position of the column corresponding to the maximum suppression of
the barrier.}
\label{Fig-CurrL}
\end{figure}

As we have already mentioned, the penetration of a pancake into an
irradiated superconductor happens in two steps: hopping from the
surface to a column, and detachment from the column into the bulk.
The resulting effective barrier is equal to the maximum of the two
barriers for the two processes.  In addition, the nucleation of a
vortex at the column (the first step) can occur via two channels:
(i) as motion of a vortex from the surface to a column or (ii) by
nucleation at and the subsequent motion of an {\it antivortex}
from the column to the surface.  The effective barrier for this
first step is, thus, the smaller one: the process goes through the
easier channel.  For every fixed position of the column there
exists a certain value, $j_{cd}(l)$, of the surface current above
which the state with the one flux quantum trapped in the column
becomes energetically favorable:
\[
j_{cd}(l)=\frac{\ln (l+\sqrt{l^{2}-1})}{\sqrt{l^{2}-1}}.
\]
This current is plotted in Fig.\ \ref{Fig-CurrL}.  The magnetic
field corresponding to this current is given by
\[
H_{cd}(L)=\frac{\Phi _{0}}{4\pi \lambda \sqrt{L^{2}-R^{2}}}\ln \left(
\frac{
L+\sqrt{L^{2}-R^{2}}}{R}\right) .
\]

Using the total energy of the vortex located at the point $(x,0)$,
\[
U_{v}(x)=U_{int}(x)+\phi _{j}(x),
\]
one can calculate the activation barrier for the vortex to hop
from a surface to the column (process 1 in Fig.\
\ref{Fig-Processes}),
\[
{\cal U}_{1}(l,j,R/\xi)=\max_{0<x<l-1}U_{v}(x).
\]
Analogously, making use of the expression for the energy of the
couple, the antivortex at the point $(x,0)$ and the vortex inside
the column,
\[
U_{av}(x)=U_{int}^{(av)}(x)-\phi _{j}(x)-j\sqrt{l^{2}-1},
\]
one obtains the barrier for the motion of the antivortex from the
column to the surface (process 2 in Fig.\ \ref{Fig-Processes}),
\[
{\cal U}_{2}(l,j,R/\xi)=\max_{0<x<l-1}U_{av}(x).
\]
The trajectory of the process, i.e. the channel that will actually
govern vortex penetration, depends on the magnitude of the applied
current.  At small currents the surface-to-column process
dominates while at sufficiently large currents the
vortex-antivortex mechanism comes into play.  The characteristic
current, $j_{av}(l)$, separating these two regimes depends on the
position of the column and is determined by the solution of the
equation ${\cal U}_{1}(l,j_{av})= {\cal U}_{2}(l,j_{av})$.  The
plot of $j_{av}(l)$ is shown in Fig.\ \ref{Fig-CurrL} together
with $j_{cd}(l)$.  The barrier to activate the vortex from the
column into the bulk of superconductor (process 3 in Fig.\
\ref{Fig-Processes}) is given by
\[
{\cal U}_{3}(l,j,R/\xi)\! =\! \left\{
\begin{array}{l}
\!\max_{x>l+1}\left[ U_{v}(x)\right] , j<j_{cd}(l) \\
\!\max_{x>l+1}\left[ U_{v}(x)\right]\! -\!U_{tr}(l,j), j\!>\!j_{cd}(l)
\end{array}\right|.
\]
Finally, the total barrier corresponding the channel
surface$\to$column$\to$bulk reads:
\begin{equation}
{\cal U}(l,j,R/\xi)=\max \left[ \min \left( {\cal U}_{1},{\cal
U}_{2}\right) ,{\cal U}_{3}\right] . \label{FullBarr}
\end{equation}
In the following we will calculate the effective reduction of the
surface barrier by the column
\begin{equation}
\delta {\cal U}(l,j)={\cal U}(l,j,R/\xi)- {\cal U} _{0}(j,R/\xi),
\end{equation}
which does not depend on the ratio $R/\xi$. Here
\begin{equation}
    {\cal U} _{0}(j,R/\xi)=\ln \frac{1.47R}{\xi }
    +\ln\frac{2}{j}-1,
\end{equation}
is the barrier for pancake penetration through an ideal
surface.\cite{Kopylov90,Mints93,Burlachkov94,NoteIdSurf}
\begin{figure}
\epsfxsize=3.2in \epsffile{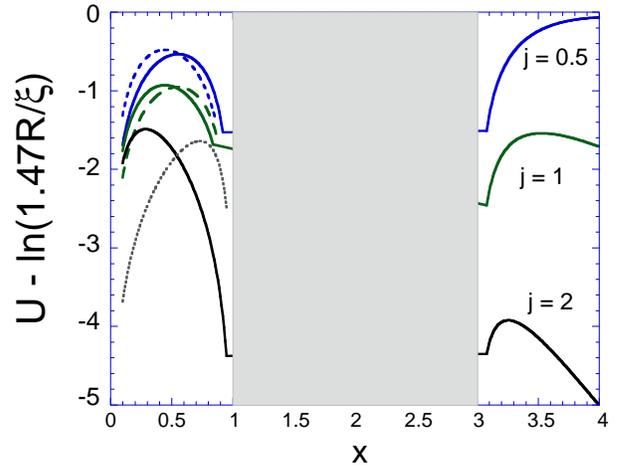}
\caption{The evolution of the energy profiles with increasing
screening surface current $j$ for $l=2$. On the lefthand side:
solid lines represent the energy profiles, $U_{v}(x)$, for the
vortex moving from the surface to the column, and dashed lines
represent the energy profiles for an antivortex, $U_{av}(x)$,
moving in the opposite direction. } \label{Fig-En_x}
\end{figure}
\begin{figure}
\epsfxsize=3.2in \epsffile{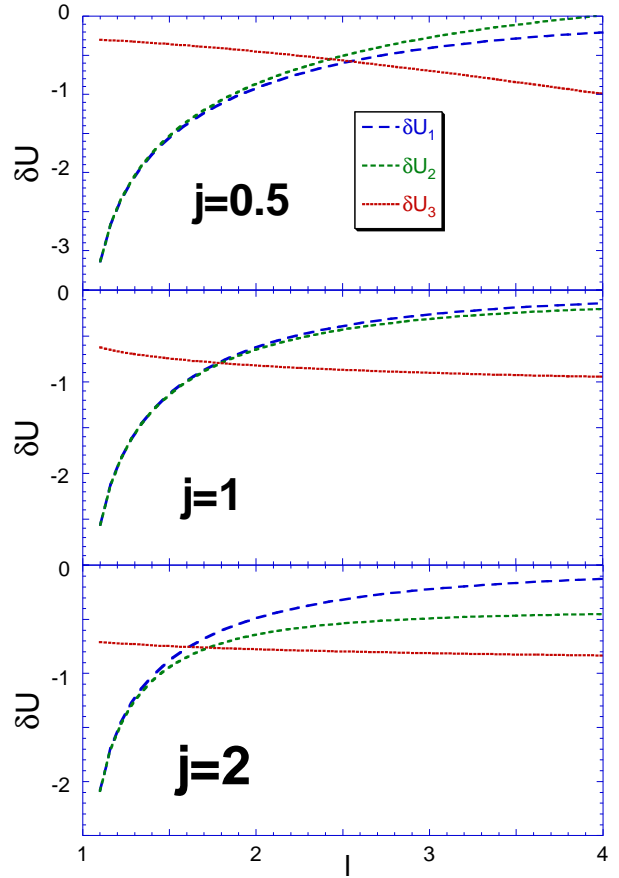} \caption{Dependence of the
energy barriers on relative distance from the surface $l=L/R$ for
the three processes shown in Fig.\ \protect\ref{Fig-Processes} for
several values of $j$.} \label{Fig-Barrl}
\end{figure}

\begin{figure}
\epsfxsize=3.in \epsffile{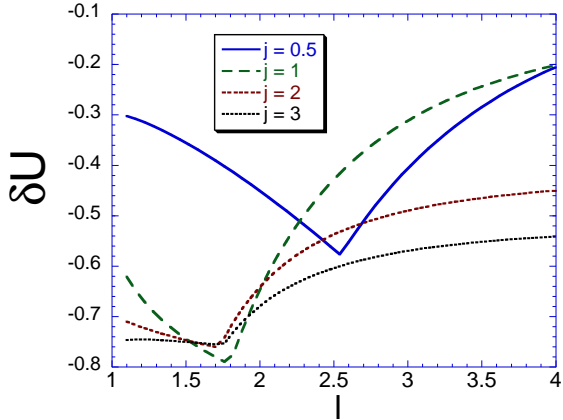} \caption{ l-dependence of the
total barrier for penetration from the surface into the bulk for
several values of $j$.} \label{Fig-totBarr}
\end{figure}
\begin{figure}
\epsfxsize=2.5in \epsffile{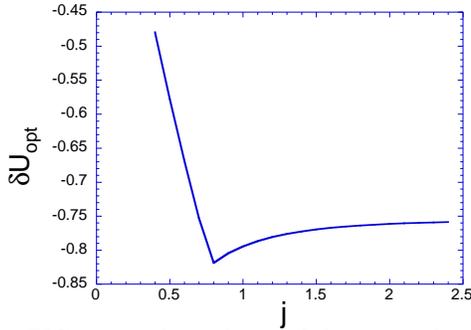} \caption{ $j$- dependence
of the energy barrier reduction for penetration of a vortex from
the surface into the bulk for the \emph{optimum} column location,
i.e. for the minimum positions of $\delta U(l,j)$ plotted in Fig.\
\protect\ref{Fig-totBarr}. One can expect that this plot
determines the reduction of the surface barrier in real irradiated
samples in the single column regime.} \label{Fig-optBarr}
\end{figure}
We will now turn to a numerical evaluation of the barriers.  Fig.\
\ref{Fig-En_x} illustrates the evolution of the energy profiles
$U_{v}(x)$ and $U_{av}(x)$ with increasing current for $l=2$.  As
one can see, at small current the position of the maximum energy
is located at the righthand side of the column while at large
current it is located between the surface and column.  Fig.\
\ref{Fig-Barrl} shows examples of the $l$ dependence of the
column-induced barrier changes $\delta {\cal U}_{1}(l,j)$,
$\delta {\cal U}_{2}(l,j)$ and $\delta {\cal U}_{3}(l,j)$ for
different $j$.  The barriers $ {\cal U}_{1}(l,j)$ and ${\cal
U}_{2}(l,j)$ are rapidly suppressed when the column approaches the
surface, mainly because of the divergence of the surface
current.  On the other hand, the barrier $\delta {\cal
U}_{3}(l,j)$ increases slowly with the decrease of $l$. Thus, for
large $l$ the total barrier is determined by ${\cal U}_{1}(l,j)$
(or ${\cal U}_{2}(l,j)$) and for small $l$ it is determined by
${\cal U}_{3}(l,j)$. Fig.\ \ref{Fig-totBarr} shows the
$l$-dependence of the total barrier ${\cal U}(l,j)$ for different
$l$.  For every value of the surface current (external field)
there is an optimum location of the column $l_{opt}$ for which
suppression of the barrier is maximum.  It corresponds to a
transition between the two mechanisms of penetration, i.e., ${\cal
U}_{3}(l_{opt},j)=\min \left( {\cal U}_{1}(l_{opt},j),{\cal
U}_{2}(l_{opt},j)\right) $. The line $l_{opt}(j)$ is shown at
Fig.\ \ref{Fig-CurrL} together with $j_{cd}(l)$ and $j_{av}(l)$.
Fig.\ \ref{Fig-optBarr} shows the current dependence of the
barrier suppression for an optimally located column $\delta {\cal
U}_{opt}(j)\equiv \delta {\cal U}(l_{opt},j)$.  This plot
represents the main result of the paper.  The important feature of
this dependence is that at high currents it saturates at $\approx
-0.76$.

An important manifestation of the surface barrier in high-T$_{c}$
superconductors is the enhancement of the effective penetration
field.\cite{Kopylov90} When the penetration field is limited by
thermal penetration through the surface barrier, the effective
penetration field is determined by the equation ${\cal
U}_{0}(H_{p0})=CT$, where the numerical constant $C\approx 20-40$
is determined be the experimental time scale and by the attempt
frequency.\cite{Kopylov90,Burlachkov94} This gives the exponential
temperature dependence of the penetration field. In irradiated
samples the penetration field is expected to be shifted to lower
value, which is determined by ${\cal U}_{0}(H_{p})+\delta {\cal
U}(H_{p})=CT$.  In the field range $ H_{p}\gg \frac{\Phi
_{0}}{4\pi \lambda R}$ where the barrier reduction approaches the
constant value $-0.76\;s\varepsilon _{0}$ we arrive at the very
simple result $H_{p}\approx 0.47H_{p0}$.

\section{Collective suppression of the surface barrier by column
clusters}
\label{Sec-Coll}

In the previous sections we considered suppression of the surface
barrier due to an isolated columnar defect near the surface.  Such a
mechanism does not give a full picture.  The barrier may be suppressed
even more substantially in small number of places where several columns
occur to be close to the surface.  The resulting enhancement of the
average penetration rate comes as a trade between the suppression of the
barrier and small probability of such event.  In this section we
estimate the contribution to the penetration rate and the reduction of the
effective penetration field caused by clusters of defects.
%
%
The optimal gate for the vortex entry will appear as several columns line
up next to  each other forming thus a cut thrusting the sample
normal to the surface.  The width of the cut is $2R$ and its length is
$2NR$, where $N$ is the number of columns in such a cluster.  In order
to find the entry energy due  to such a cluster we will follow the
Refs.\ \onlinecite{BuzdPhysC98,MelnPhysC01}, where the current and
field distributions near the edge of the wedge-like surface crack
where calculated.  Consider thus a cut normal to the surface made
of the chain of $N$ columns. According to Refs.\
\onlinecite{BuzdPhysC98,MelnPhysC01} the current near the tip of a
thin crack grows as $j\sqrt{2NR/x}$ when distance $x$ from the tip
of the crack (the last column) falls into interval $R<x\ll 2NR$,
and saturates at $j\sqrt{2N}$ for $x\lesssim R$.  Here $j$ is the
Meissner current far away from the crack.  The suppression of the
barrier depends on the relation between the position of the energy
maximum $x_{0}$ and $R$.
\begin{figure}
\epsfxsize=3.2in \epsffile{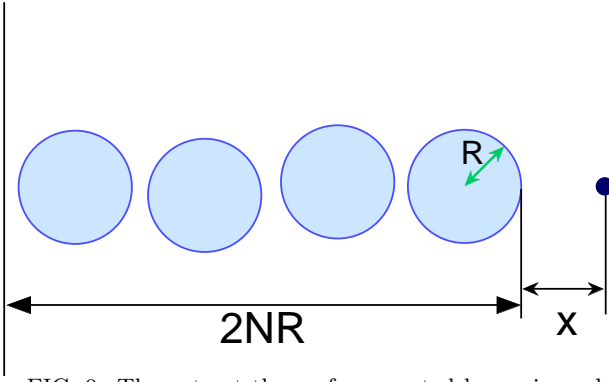} \caption{The gate at the surface
created by an improbable event when $N$ columns form a chain.  Such
gates dominate flux penetration at low temperatures.}
\label{Fig-chain}
\end{figure}

Consider the case $x_{0}>R$ first.  Making use of the results of
Refs.\ \onlinecite{BuzdPhysC98,MelnPhysC01}, we write the energy
of vortex at distance $x$ from the tip of the crack as
\cite{PhysicalUnits}
\[
E(x)=-\frac{s\Phi _{0}}{c}j\sqrt{2NRx}+s\varepsilon _{0}\ln \frac{x}{\xi }.
\]
Accordingly, the barrier for this gate configuration (the maximum
value of $E(x)$) is
\begin{equation}
    {\cal U} ={\cal U}_{0}-s\varepsilon _{0} \ln \frac{\Phi _{0}jNR}{
c\varepsilon _{0}},
\end{equation}
where the bare barrier ${\cal U}_{0}$ in real units is given
${\cal U}_{0}=s\varepsilon _{0}\ln (j_{dp}/j)$ and $j_{dp}$ is the
depairing current.  The energy $E(x)$ achieves its maximum value
at the point $ x_{0}=2\left(c\varepsilon _{0}/{\Phi _{0}j}\right)
^{2}/(NR)$, which has to satisfy conditions $R<x_{0}<NR$.

In order to estimate the effective penetration rate one has to
determine the size of the clusters giving maximal contribution to
vortex entry. The probability to find a chain of $N$ columns in a
row scales as $(\pi R^{2}n_{d})^{N}$ with $n_{d}$ being the column
concentration.  Therefore the contribution from such chains to the
penetration rate is given by
\begin{eqnarray}
\nonumber \sum_{N}(\pi R^{2}n_{d})^{N}\exp \left( \frac{s\varepsilon _{0}}{T}\ln \frac{
\Phi _{0}jNR}{c\varepsilon _{0}}\right) =\\ \sum_{N}\exp \left( -N\ln \frac{1}{
\pi R^{2}n_{d}}+\frac{s\varepsilon _{0}}{T}\ln \frac{\Phi _{0}jNR}{
c\varepsilon _{0}}\right),
\end{eqnarray}
and becomes maximal at the optimal $N$
\[
N=\frac{s\varepsilon _{0}}{T\ln(1/\pi R^{2}n_{d})}>\max \left[ 1,
\frac{2c\varepsilon _{0}}{\Phi _{0}jR}\right].
\]
Now the condition $x_{0}>R$ reduces to
\[
T>\frac{s\varepsilon _{0}}{\ln(1/\pi R^{2}n_{d})}\left( \frac{\Phi
_{0}jR}{2c\varepsilon _{0}}\right) ^{2}.
\]
In this ``high-temperature'' regime the change of the surface
barrier is:
\begin{eqnarray*}
\delta {\cal U}&\approx&-s\varepsilon _{0} \ln \frac{s\Phi _{0}jR}{2cT
\ln(1/\pi R^{2}n_{d})}\\
&=&- s\varepsilon _{0} \ln \frac{sR\Phi
_{0}H}{8\pi \lambda T\ln (1/\pi R^{2}n_{d})}
\end{eqnarray*}
where, again, $j=cH/4\pi \lambda $.

The effective penetration barrier is determined by the relation:
\begin{equation}
s\varepsilon _{0}\ln \frac{H_{c}H_{T}}{H^{2}} =CT
\end{equation}
with $H_{T}= \frac{sR\Phi _{0}}{8\pi \lambda T\ln (1/\pi
R^{2}n_{d})} $, which gives
\begin{equation}
H_{p} =\sqrt{H_{c}H_{T}}\exp \left( -\frac{2CT}{s\varepsilon _{0}}\right)
\label{HighTHp}
\end{equation}

Consider now the regime $x_{0}<R$, corresponding to low temperatures. In
this case the energy of the vortex is given by
\[
E(x)=-\frac{s\Phi _{0}}{c}j\sqrt{2N}x+s\varepsilon _{0}\ln \frac{x}{\xi }.
\]
The position of the energy maximum and the barrier are given by
\begin{eqnarray*}
x_{0} &=&\frac{c\varepsilon _{0}}{\Phi _{0}j\sqrt{2N}} \\
U &=&s\varepsilon _{0}\ln \frac{c\varepsilon _{0}}{\Phi _{0}\xi j\sqrt{2N}}
\approx U_{0}-\frac{s\varepsilon _{0}}{2}\ln N
\end{eqnarray*}
Following the same route one arrives at the optimal cluster length as
\[
N=\frac{s\varepsilon _{0}}{2T\ln (1/ \pi R^{2}n_{d})}>1,
\]
and, accordingly, the change of the barrier is
\[
\delta U=-\frac{s\varepsilon _{0} }{2}\ln \frac{s\varepsilon _{0}}{2T
\ln(1/ \pi R^{2}n_{d})}
\]
Contrary to the case of individual columns, the cluster gate
exhibits the temperature dependence: the depression of the barrier
{\it decreases} logarithmically with growing temperature.  The
change of the effective penetration field is determined by
\[
s\varepsilon _{0}\ln \frac{H_{c}}{H_{p}}-\frac{s\varepsilon _{0}}{2}\ln
\frac{s\varepsilon _{0}}{2T\ln (1/ \pi R^{2}n_{d})} =CT \\
\]
or
\begin{equation}
H_{p}=\sqrt{\frac{2T\ln (1/ \pi R^{2}n_{d})}{s\varepsilon _{0}}}H_{p0}
\label{LowTHp}
\end{equation}
This formula gives the effective penetration fields at low temperatures.
It is probably more relevant to experimental situation than the
``high-temperature'' result (\ref{HighTHp}).  We see that long clusters
of columnar defects forming the crack-like configurations further (as
compared to the effect of an individual column) suppress the surface barrier
and serve as a very effective vortex gates into the sample.  The
collective mechanism wins over the single-column mechanism at low
temperatures.  Comparing Eq.\ (\ref{LowTHp}) with the single-column result
we conclude that the at high magnetic fields the crossover between the
single column and collective regimes is expected at the crossover
temperature $T_{cr}$
\begin{equation}
T_{cr}\approx \frac{s\varepsilon _{0}}{8\ln (1/ \pi R^{2}n_{d})}
\label{Tcr}
\end{equation}
For the typical parameters of the compound Bi$_{2}$Sr$_{2}$CaCu$_{2}$O$_{x}$
at $n_{d}= 5\cdot 10^{10}$ cm$^{-2}$ (corresponding to the matching field of 1
tesla) and $R=35\AA$ the crossover is expected at $\approx$ 20 K.

Our calculations suggest that in the regime of the thermally-activated
pancake penetration through the surface barrier, the penetration field
is roughly two times smaller than the penetration field of unirradiated
samples at temperatures above the crossover temperature (\ref{Tcr}) and
decreases according to Eq.\ (\ref{LowTHp}) at lower temperatures.  This
is consistent with the recent experiment.  \cite{Gregory01}

We have investigated the influence of columnar defects in layered
superconductors on the thermally activated penetration of pancake
vortices through the surface barrier.  Depending on the position
of an isolated column the effective barrier is determined either
by the vortex hopping from the surface to the column or by the
detachment of the vortex from the column to the bulk.  For a given
external field there exists an optimum location of the column for
which the barriers the both stages are equal and the reduction of
the effective penetration barrier is maximal.  Formation of long
clusters of columnar defects thrusting the surface offers the most
convenient gates for the vortex entry, the effect of
cluster-induced depression of the surface barrier decreasing with
temperature.  Penetration through the clusters always dominates at
low temperatures. It would be very interesting to investigate
experimentally temperature dependencies of the surface barriers
and penetration rates in order to reveal the role of clusters.
The proposed mechanism of the disorder-assisted surface creep is
very general and can be extended to the point disorder containing
samples.

We thank C.\ J.\ van der Beek, M.\ Konczykowski, and A.\ Mel'nikov
for helpful discussions.  This work was supported by the U.S. DOE,
Office of Science, under contract \# W-31-109-ENG-38.

\end{multicols}

\begin{references}
\bibitem{NoteIdSurf} Note that the column radius $R$ appears in this
equation only because our unit of current contains $R$.  In real
units this barrier, of course, does not depend on $R$.
\bibitem{PhysicalUnits} Because we only have qualitative estimates in
this section, we use physical units from the very beginning.
\end{references}
\end{document}